\documentstyle[aps,psfig]{revtex}
\twocolumn
\tighten
\begin{document}
\draft
\preprint{TPI-MINN-01/41 $\;\;$
          UMN-TH-2023-01 $\;\;$
          %ITP-SB-00-20   $\;\;
}

\newcommand{\nc}{\newcommand}
\nc{\vivi}{very interesting and very important}
\nc{\lnc}{\Lambda_{NC}}
\nc{\tmf}{\theta_{\mu5}}
\nc{\al}{\alpha}
\nc{\ga}{\gamma}
\nc{\de}{\delta}
\nc{\ep}{\epsilon}
\nc{\ze}{\zeta}
\nc{\et}{\eta}
\newcommand{\th}{\theta}
\nc{\tmn}{\theta_{\mu\nu}}
\nc{\Th}{\Theta}
\nc{\ka}{\kappa}
\nc{\la}{\lambda}
\nc{\rh}{\rho}
\nc{\si}{\sigma}
\nc{\ta}{\tau}
\nc{\up}{\upsilon}
\nc{\ph}{\phi}
\nc{\ch}{\chi}
\nc{\ps}{\psi}
\nc{\om}{\omega}
\nc{\Ga}{\Gamma}
\nc{\De}{\Delta}
\nc{\La}{\Lambda}
\nc{\Si}{\Sigma}
\nc{\Up}{\Upsilon}
\nc{\Ph}{\Phi}
\nc{\Ps}{\Psi}
\nc{\Om}{\Omega}
\nc{\ptl}{\partial}
\nc{\del}{\nabla}
\nc{\be}{\begin{equation}}
\nc{\ee}{\end{equation}}
\nc{\bea}{\begin{eqnarray}}
\nc{\eea}{\end{eqnarray}}
\nc{\ov}{\overline}
\nc{\gsl}{\!\not}
\newcommand{\s}{\mbox{$\sigma$}}
\newcommand{\bi}[1]{\bibitem{#1}}
\newcommand{\fr}[2]{\frac{#1}{#2}}
\newcommand{\gm}{\mbox{$\gamma_{\mu}$}}
\newcommand{\gn}{\mbox{$\gamma_{\nu}$}}
\newcommand{\Le}{\mbox{$\fr{1+\gamma_5}{2}$}}
\newcommand{\R}{\mbox{$\fr{1-\gamma_5}{2}$}}
\newcommand{\GD}{\mbox{$\tilde{G}$}}
\newcommand{\gf}{\mbox{$\gamma_{5}$}}
\newcommand{\Ima}{\mbox{Im}}
\newcommand{\Rea}{\mbox{Re}}
\newcommand{\Tr}{\mbox{Tr}}
\newcommand{\psl}{\slash{\!\!\!p}}
\newcommand{\cp}{\;\;\slash{\!\!\!\!\!\!\rm CP}}
\newcommand{\qq}{\langle \ov{q}q\rangle}
\def\ga{\mathrel{\raise.3ex\hbox{$>$\kern-.75em\lower1ex\hbox{$\sim$}}}}
\def\la{\mathrel{\raise.3ex\hbox{$<$\kern-.75em\lower1ex\hbox{$\sim$}}}}

%@twocolumnfalse\endcsname  

\title{Breaking CPT by mixed non-commutativity}

\vspace{1cm}

\author{Irina Mocioiu$^1$\footnote{mocioiu@insti.physics.sunysb.edu},
Maxim Pospelov$^2$\footnote{pospelov@mnhepw.hep.umn.edu} 
          and Radu Roiban$^{1}$\footnote{
roiban@insti.physics.sunysb.edu}}

\vspace{1cm}

\address{$^1$ C.N. Yang Institute for Theoretical Physics\\
State University of New York, Stony Brook, NY 11794-3840\\ $\;$\\
$^2$ Theoretical Physics Institute, School of Physics and Astronomy \\
         University of Minnesota, 116 Church St., Minneapolis, MN
         55455, USA
         }
\date{\today}

\maketitle

\begin{abstract}
The mixed component of the non-commutative parameter $\theta_{\mu M}$, 
where $\mu = 0,1,2,3$ and $M$ is an extra dimensional index may violate 
four-dimensional CPT invariance. We calculate one and two-loop induced 
couplings of $\theta _{\mu 5 }$ with the four-dimensional 
axial vector current and with the CPT odd dim=6 
operators starting from five-dimensional 
Yukawa and $U(1)$ theories. The resulting bounds from clock comparison 
experiments place a stringent constraint on $\theta _{\mu 5 }$, 
$|\theta_{\mu 5 }|^{-1/2}\mathrel{\raise.3ex\hbox{
$>$\kern-.75em\lower1ex%
\hbox{$\sim$}}}5\times 10^{11}$ GeV. Orbifold projection and/or
localization of fermions on a 3-brane lead to CPT-conserving
physics, in which case the constraints on $\tmf$ are softened. 

\end{abstract}
 
%\vspace{4cm}

%\hfill\eject

\section{Introduction}

Non-commutative field theories and their realizations in string theory 
have been a subject of intensive theoretical research over the past few 
years (See, e.g. \cite{SW,DN}). Much of this excitement has gone totally 
unnoticed by particle phenomenology for the following simple reason. 
The presence of the antisymmetric tensor $\theta_{\mu\nu}$ 
as a constant background violates Lorentz invariance, a possibility 
excluded to an impressive accuracy by various 
low-energy precision measurements (See, e.g. \cite{Kost}). 
In our previous paper,
Ref. \cite{MPR}, we have shown that in the low-energy effective interaction 
linearized in the non-commutative parameter, $\theta_{\mu\nu}$ couples to 
the nucleon spin, $\bar N \sigma_{\mu\nu} N$ with the strength 
proportional to the cube of the characteristic hadronic scale, $
\Lambda_{\rm hadr} \sim $1 GeV. 
This analysis has been done at the tree level in order to avoid potential 
problems with the issues of renormalizability of the non-commutative 
theories, the necessity to introduce a cutoff, etc. This coupling 
generates an additional, magnetic-field-independent, contribution to the
nucleon Larmor frequency. Therefore, this interaction has the signature of a 
constant magnetic field of a fixed direction and can be searched through 
a precise monitoring of siderial variation of the magnetic field. 
Ref. \cite{Amh} places the limit on the possible size of such an interaction 
at the level of $\nu$=100 nHz. Comparing it with the result of the theoretical 
calculation in \cite{MPR}, one arrives at an incredibly strong constraint,
$\lnc = 1/\sqrt{\theta }\mathrel{\raise.3ex\hbox{
$>$\kern-.75em\lower1ex%
\hbox{$\sim$}}}5~\times 10^{14}$ GeV. If non-commutativity is realized 
somehow
only in 
the leptonic sector, $\theta_{\mu\nu}\bar N \sigma_{\mu\nu} N$ interaction 
is still generated, although with $(\alpha/\pi)^2$ suppression compared to the
non-commutative QCD case. This relaxes the limit down to 
$ 10^{12}$ GeV level. Later it was argued in Refs. \cite{UCSC,Carone} that
$\sigma_{\mu\nu}\theta_{\mu\nu}$ operator can be generated at the loop  level 
with quadratically divergent integral which may bring even tighter bounds
which scale with the cutoff.

Subsequent analyses have addressed the possibility to observe $\theta_{\mu\nu}$
in future collider experiments \cite{Hew}, and in the neutral $K$ and $B$
meson systems \cite{Hinch}. If the former cannot do much better than 
$\Lambda_{NC} \sim 1$ TeV, the neutral kaons could in principle 
be quite sensitive to $\lnc$. However, even with the most favorable 
assumptions that the non-commutativity generates somehow an effective
$\Delta F = 2$ transition $\theta (\bar d O s) (\bar d O s)$ of unsuppressed 
strength, one can get only to the level $\lnc \sim 10^9$ GeV. This is in the 
range already excluded by the clock comparison experiments! Generically, any 
attempt to construct a fully non-commutative Standard Model (e.g. Ref. 
\cite{NCSM}) will have to comply with the bound obtained in \cite{MPR}
which would make $\theta_{\mu\nu}$ totally unobservable for 
conventional particle physics experiments. 
(A possible exception could be measurements of the refraction index over 
cosmological distances \cite{CFJ} or the search for the
imprints of non-commutative inflation
\cite{Greene}, when the characteristic energy scales could 
be quite high and compensate for the extreme smallness of $\lnc^{-2}$). Other
relevant works on the observational consequences of non-commutativity 
include \cite{massgrave}.

In this paper we study phenomenological consequences of mixed 
non-commutativity $\theta_{\mu M}$, where $\mu$ is a normal 4-d index and 
$M$ is along extra dimensions. The presence
of such a component is perceived by a four-dimensional observer as a 
constant 4-vector, which obviously breaks Lorentz invariance.
The purpose of this work is to show that in certain classes 
of models this background may also break 4-dimensional CPT invariance. 
Indeed, the transformation of $\theta_{\mu M}$ under the CPT reflection is 
similar to the behavior of charge times the U(1) field strength,
$e F_{\mu M}$, for which we know that $CPT(e F_{\mu M}) =
CPT(e \partial/\partial x^M A_\mu)=-1$. Note that parity is defined here in 
a 4-dimensional sense and thus $P(\partial/\partial x^M)=1$.

Unlike the previous case with the breaking of Lorentz invariance by 
$\theta_{\mu\nu}$ for which no plausible low-energy physics 
motivation exists, one can think of 
baryogenesis-motivated reasons to study the CPT non-invariant interactions. 
Namely, we refer to an interesting idea \cite{Kuzmin} that the breaking of 
CPT effectively comprises two out of three Sakharov's conditions needed
for baryogenesis, requiring only the breaking of the baryon number as an
extra ingredient. If baryogenesis happens at the early cosmological
epoch with characteristic energy scales $E^4 \sim 1/(\theta_{\mu M})^2$, 
a dramatic power-like suppression by $\lnc$ may be lifted. 

We study the possibility of CPT violation by $\theta_{\mu5}$ 
and find that it is possible in the 5-dimensional model 
with Dirac fermions and scalar 
particles and/or gauge bosons living in all 5 dimensions. After the 
compactification 
of the fifth dimension, the zero level Kaluza-Klein mode of the fermion 
acquires the coupling with $\tmf$ through the axial-vector current
or the CPT non-invariant dim=6 operators. 
Specifying this to the case of QED, we effectively 
get 
%$\tmf \bar e \gamma_\mu \gamma_5 e $,  $\tmf \bar N\gamma_\mu \gamma_5 N$ or 
$\tmf \partial_\mu F_{\alpha\beta}\bar N\sigma^{\alpha\beta} \gamma_5 N$
which results in the strong bounds on $\tmf$. 
On the other hand, we show that
in the models with intrinsically 4-d fermions, e.g. localized on a 
3+1-dimensional domain wall, the CPT may be conserved and only 
even powers of $\tmf$ may appear in the low-energy effective lagrangian.

\section{5-d model with low-energy 4-d CPT violation}

We begin by reminding that the nonvanishing commutation relations 
\be
[{\hat x}^a,\,{\hat x}^b]=i\theta^{ab}~~~~~~~~
[\theta^{ab},\,{\hat x}^c]=0
\ee
in the coordinate space ($ a,b,c = 0,1,2,3,M,..N $ ) lead to the modification 
of the interaction terms in the field theory through 
the Moyal product, given by
\be
\phi_1*\phi_2 (x)=e^{i\frac{1}{2}\theta^{\mu\nu}\frac{\partial}
{\partial\xi^\mu}
\frac{\partial}{\partial\zeta^\nu}} \phi_1(x+\xi)\phi_2(x+\zeta)
|_{\xi=\zeta=0}.
\label{eq:starprod}
\ee
Working in the framework of an effective field theory, and assuming
smallness of $E^2\theta$, where $E$ is the characteristic energy scale in
the problem, we linearize the $*$-product to get a combination 
\be
\phi_1*\phi_2 (x) = \phi_1\phi_2 (x) +
i\frac{1}{2}\theta^{\mu\nu}
\frac{\partial}{\partial x^\mu}\phi_1
\frac{\partial}{\partial x^\nu}\phi_2 +...
\label{reduction}
\ee

Specializing the second term in this expansion to the case of 
$\theta_{\mu5}$, we immediately observe the presence of $\partial/\partial y$
derivative along the fifth dimension. We take this dimension to be 
compact with the radius smaller than $10^{-17}$ cm so that 
the Kaluza Klein modes of fermions, gauge bosons and scalars are 
heavy enough to avoid particle physics constraints. 
Our goal is to integrate out these heavy states and 
obtain the low-energy effective action for the zero-level KK modes. 
Lowest energy modes do not have any $y$-dependence. Therefore, $\tmf$
may only appear in the loop-induced amplitudes with excited KK modes inside.
Moreover, since the term we are interested in contains 
$i\partial/\partial y = p_5$ in one of the vertices, 
we have to find an amplitude that would contain an odd function of 
$p_5$ in the propagators because otherwise the loop amplitude will 
vanish upon summation of positive and negative $p_5$ modes. An 
odd function of $p_5$ may come only from the fermion 
propagators. Therefore, in order to get a non-vanishing result linear in 
$\theta_{\mu M}$ we have 
to allow fermions to propagate in a full five-dimensional space.

As a warm-up exercise we consider a
two-loop-gen- erated 
$\bar \psi \sigma_{MN}\theta^{MN} \psi $ amplitude 
in the non-commutative scalar-fermion theory with Yukawa interaction
$\lambda \bar \psi \psi \phi + i\lambda \th_{MN}  
\bar \psi \partial_N \psi \partial_M\phi$. This calculation is similar
to the 4-dimensional calculation performed in Ref. \cite{UCSC}. The $\mu 5$
component of $\theta$ couples to $\bar \psi \sigma_{\mu, N=5} \psi =
-\bar \psi \gamma_\mu \gamma_5 \psi$, which is a CPT-odd operator in 
four dimensions. Taking into account the contribution of the first excited 
KK level, we arrive at the following two-loop induced amplitude,
\be
\tmf (\bar\psi \gamma_\mu\gamma_5 \psi )\times {\lambda^4 m
\over 32 \pi^4} M^2 \ln(\Lambda_{UV}^2/M^2).
\label{yukawa}
\ee 
In this formula $m$ and $M$ are the masses of the zeroth and 
first KK levels and $ \Lambda_{UV}$ is an ultraviolet cutoff. 
The summation over the Kaluza Klein tower diverges as $N^4$.
Thus we would have to cut it at some value $N_{\rm max}^4$ 
which exactly corresponds to an initial $\Lambda^4_{UV}$ 
divergence of the 5-dimensional two-loop integral. 

We would like to comment at this point that in a 
supersymmetric four-dimensional 
theory there can be a natural scale for the ultraviolet cutoff
related to the soft breaking masses, 
$\Lambda_{UV}\sim m_{\rm soft}$. We base this argument on the similarity of
$\theta_{MN}\bar \psi \sigma_{MN} \psi$ operator 
with the fermion anomalous magnetic moment 
which is not supersymmetrizable and thus must vanish in the exact SUSY limit
\cite{FR}.
%$\theta^{\mu\nu} \bar\psi \sigma_{\mu\nu} \psi$ operator is not 
%supersymmetrizable in the effective action
%\footnote{This can be easily  seen by noting that
%in superfield notation it is written as 
%$\theta^{{\dot\alpha}{\dot\beta}}{\bar D}_{\dot\alpha}{\bar \Phi}_1
%{\bar D}_{\dot\beta}{\bar \Phi}_2+h.c.$ where 
%$\theta^{{\dot\alpha}{\dot\beta}}$ is the anti-self-dual 
%%part of $\theta^{\mu\nu}$.
%But, because of the symmetry of $\theta^{{\dot\alpha}{\dot\beta}}$, 
%this is not integrable
%to a modified chiral superspace measure.} 
%\cite{FR} and thus must vanish in the exact SUSY limit. 
We believe, however, that the explicit calculation of 
$\theta_{MN}\bar \psi \sigma_{MN} \psi$ interaction 
in softly broken supersymmetric theory is needed to clarify this matter. 
To stay on the conservative side, we shall assume that the cutoff
is not higher than few hundred GeV. Having the answer (\ref{yukawa}) at hand,
one can try to determine the level of sensitivity to $\tmf$ 
of different CPT and Lorentz violation searching experiments 
\cite{Amh,Trap,Heck,Harvard1,Harvard2} by identifying $\phi$ with 
the Higgs boson and $\lambda$ with the light fermion Yukawa coupling. 
Clearly, one would have a serious suppression because all the relevant Yukawa 
couplings are quite small.

Thus, we are bound to explore gauge theories and we limit our discussion 
to the case of the five-dimensional non-commutative $U(1)$. 
One would naively guess that with the use of the 
non-commutative QCD in the bulk, 
the effective interaction is going to be at least two orders of 
magnitude stronger, simply because $\alpha_s \gg \alpha$. 
In fact QED and QCD may produce comparable results as it is known that 
multiple KK states accelerate the renormalization group running and 
already after few thresholds are taken into account, $\alpha_s \simeq \alpha$ 
\cite{DDG}.

First, we note that $m \theta_{MN}{\bar\psi}
\sigma_{MN}\psi$ is not generated at one-loop 
level. We also disagree with the claim of the Ref. 
\cite{Carone} that $\theta_{MN}{\bar\psi} \sigma_{MN}\psi$ 
interaction can 
be generated at one-loop level in a non-commutative $SU(N)$ 
theory simply because the amplitude, calculated in  
\cite{Carone}, is proportional to $\psl - m $ and therefore vanishes on-shell. 
In what follows we restrict our calculation to the one-loop level and 
compute the effective dimension 6 operators. Clearly, the use of
higher dimensional operator will generically be suppressed by, say,
$\Lambda_{\rm hadr}^2/\Lambda_{UV}^2$ compared to the leading 
dimension operator, $m {\bar \psi} \sigma_{MN}\psi$. In our case, this suppression
is not going to be dramatic since  we choose a low value for the cutoff.

%%%%%%%%%%%%%%%%%%%%%%%
\begin{figure}
 \centerline{%
   \psfig{file=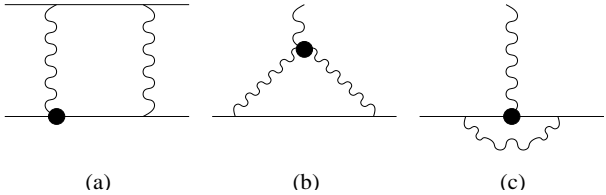,width=8cm,angle=0}%
         }
\vspace{0.1in}
 \caption{One-loop diagrams in the 5-d non-commutative $U(1)$ 
theory, which generate CPT non-invariant dim=6 operators
in four dimensions. Heavy dot represents the interaction term
with $\tmf$ which can be placed at any vertex in 
(a) non-commutative box diagram + photon permutation, 
(b) non-commutative ``bosonic'' penguin. This diagram 
does not exist in a normal, commutative $U(1)$ theory.  
(c) non-commutative ``fermionic'' penguin.
}
\end{figure}
%%%%%%%%%%%%%%%%%%%%%%%

The relevant set of diagrams, Fig. 1, contains boxes, ``bosonic'' and 
``fermionic'' penguins with all possible insertions of the $\theta$-dependent
vertices. We follow the linearized version of the Feynman rules for
the non-commutative $U(1)$ theory given, for example, in \cite{Hay}.
As in the previous example, loops contain heavy KK modes. 
We simplify our calculation by working in zeroth order in the mass 
of the external fermions. Thus, in the box diagrams we can 
neglect all external momenta after which it immediately vanishes, as 
there is only the loop momentum to be contracted with $\theta_{MN}$. 
The diagrams 1b and 1c produce the same operator. Taking into account 
the first KK mode propagating inside the loop, we get
\be
{\cal L}_{1b+1c}=\fr{5e_i^2e}{48\pi^2}\left[1+\fr{9}{10}\fr{e_i}{e}\right]
\theta_{\mu5}
(\bar\psi_i i\si^{\alpha\beta} \gamma_5\psi_i) \partial_\mu F_{\alpha\beta},
\label{4ferm}
\ee
which turns out to be independent of the KK mass $M$.   The summation 
over the whole tower is linearly divergent and gives $N_{\rm max}$. 
In the square 
brackets of eq. (\ref{4ferm}), 1 is the contribution of diagram 1b and 
$(9e_i)/(10e)$ is of 1c. 
%In models where only
%QED sector contains $\theta_{\mu\nu}$ one should retain only the contribution 
%of diagram 1b. The numerical 
%change due to inclusion or omission of diagram 1c is not crucial
%for our discussion. 

The interaction (\ref{4ferm}) obviously breaks CPT. In addition to the 
EDM-like interaction, $ F_{\alpha\beta}
(\bar\psi_i i\si^{\alpha\beta} \gamma_5 \psi_i)$, eq. (\ref{4ferm})  
has an extra derivative 
$\partial_\mu$ which changes sign under either $P$ or $T$ transformation.
In order to obtain phenomenological consequences of this interaction, 
we specialize the result (\ref{4ferm}) to the case of the light quarks.
For the matrix elements $\Delta q$ of $\bar q \sigma_{\mu\nu} \gamma_5 q$ 
operators over a nucleon, we use the results of lattice \cite{Lat} and  
QCD sum rule \cite{PR} calculations: $\Delta u_p=\Delta d_n\simeq 0.8$ 
and  $\Delta u_n=\Delta d_p\simeq -0.2$ 
Thus we get 
\be
V = \fr{5e^3N_{\rm max}}{48\pi^2}\partial_\mu F_{\alpha\beta}\tmf 
\left(\kappa_p \bar p i\sigma_{\alpha\beta}\gamma_5 p
+\kappa_n \bar n i\sigma_{\alpha\beta}\gamma_5 n \right)
\label{V}
\ee
where
\be
  \kappa_p \simeq 0.11\,;\,\,\,\kappa_n \simeq 0.08.
\ee
The use of the constituent quark model would produce $50\%$ larger results.

One can treat $\partial_\mu F_{\alpha\beta}$ in eq. (\ref{V}) 
in two different ways. The most straightforward approach would be to 
try to estimate the photon loop nucleon 
self-energy diagram with one of the vertices given by eq. (\ref{V}). 
One can easily check that the Lorentz structure of such a diagram will be 
proportional to $\tmf \bar N \gamma_{\mu}\gamma_5 N$. As to the numerical 
result of ultraviolet divergent integration, we cannot 
estimate it better than $O(e\times
{\rm loop~ factor}\times \Lambda^3)$ where $\Lambda$ can be anywhere 
between $m_\pi$ and $m_\rho$. Such a result could be considered as
an order-of-magnitude estimate at best.

We prefer to use a different method and directly calculate the nuclear matrix element
of interaction (\ref{V}). We exploit the fact that the  
gradient of the electric field inside a large nucleus 
is approximately constant,
\be
\partial_i F_{0j} \simeq \delta_{ij}\fr {Ze}{R^3} \simeq  \delta_{ij}e\fr{Z}{A}
{\rm fm}^{-3}.
\label{grad}
\ee
For a non-relativistic nucleon, inside the nucleus, $V$ reduces to the 
product of the nucleon spin operator and $\theta_{i5}$. 
The wave function of an external nucleon 
is concentrated mainly inside the nuclear radius $R$. Therefore, the 
nuclear matrix element reduces to a trivial angular part.  
Assuming one valence nucleon with orbital momentum $L$ above closed 
shells, we 
arrive at the final form of the interaction between the nuclear 
spin $I$ and the external vector $\theta_{i5}$ 
\be
V = \fr{20}{3} \fr{Z}{A}N_{\rm max}\alpha^2 \kappa_{p(n)} a_{LI} (\vec{I}\cdot 
\vec{\theta}){\rm fm}^{-3}.
\label{final}
\ee 
Here $a_{LI} = \langle \vec S \cdot \vec I\rangle_{LI} $ 
is a trivial combination of $I(I+1)$ and $L(L+1)$. $a=1$ for $L=0$. 

The two most sensitive experiments, Ref. \cite{Amh} and
Ref. \cite{Harvard1}, use $^{199}Hg$ and $^{129}Xe$ whose spins 
are carried by external neutrons. Choosing $N_{\rm max} = 1$ and 
comparing the size of $V$ with the experimental accuracy of Ref. 
\cite{Harvard1}, $|V|< 2\pi \times 100$ nHz,
we deduce the level of sensitivity to the presence of mixed non-commutativity
\be
|\theta_{i5}|  \mathrel{\raise.3ex\hbox{
$<$\kern-.75em\lower1ex%
\hbox{$\sim$}}} ( 5\cdot 10^{11} {\rm GeV}    )^{-2}
\label{finlim}
\ee
Interestingly enough, we did not gain anything in terms of $\lnc$ compared to 
our previous limit on $\theta_{ij}$ \cite{MPR}, even though in addition to Lorentz symmetry 
the effective interactions (\ref{yukawa}) and (\ref{4ferm}) violate CPT. 
This is because in the case of mixed non-commutativity, we had to resort to
a loop level and used low energy (small) value of $\alpha$. 
This limit can be improved by one or two orders 
of magnitude if one computes the two-loop level 
NCQCD induced $\th_{MN}\bar\psi \si_{MN} \psi$ interaction.

Do the results (\ref{yukawa}) and (\ref{4ferm}) mean that any  
five-dimen- sional model with 
$\tmf \neq 0$ would violate  CPT in the low-energy regime? The answer is 
no, as we can easily construct a counterexample. Indeed, let us
consider a model where all fermions stay confined to a 3+1 
domain wall and only gauge bosons live in the bulk. 
Then the parity along the extra dimension, $y\rightarrow -y$
will forbid any odd powers of coupling of $\tmf$ to the four-dimensional 
fermions. Another obvious trick which helps to get rid of CPT violation 
at low energies is the orbifold projection in 5 dimensions. The same 
argument of exact parity in $y$ coordinate will prohibit (\ref{V}). 
As the result of orbifolding and/or fermion localization, only 
quadratic couplings in $\tmf$ could be important.
Let us estimate a possible level of sensitivity to $\tmf$ in this case.

There are many four-fermion operators that may appear and 
violate Lorentz invariance.  It is easy to see, for example, 
that $c\,\tmf \theta_{\nu5} (\bar q \gamma_{\mu}\gamma_5 q)
(\bar q \gamma_\nu q)$ may arise
due to the interaction with $SU(2)$ KK gauge bosons.  The component 
$\theta_{05}\theta_{i5}$  will couple to the spin of the external 
 neutron in the mercury nucleus  with the strength proportional to 
the nuclear density, $0.17~{\rm fm}^{-3}$. Assuming again  that $c \sim 
100$ GeV$^2$ times the loop factor, we arrive at the level 
of sensitivity 
$\lnc \sim 10^7 $ GeV. It is significantly milder constraint, although 
severe enough to exclude the influence of $\tmf\theta_{\nu5}$ on any plausible 
terrestrial particle physics experiments.

\section{Conclusions}

The phenomenological consequences of the CPT violation 
are well understood, but the explicit models which break 
CPT are hard to find. In this paper we have demonstrated
that the violation of the four-dimensional CPT invariance is possible in the 
presence of mixed non-commutativity $\tmf$. In particular, this happens 
when fermions are allowed to propagate in the five-dimensional bulk. 
We have shown that the Yukawa or gauge interaction 
of these fermions generate  an effective four-dimensional 
$\tmf \bar \psi \gamma_\mu \gamma_5 \psi $ as well as higher dimensional
CPT-violating operators. 
Of course, an important ingredient 
in this picture is an indefinite parity for a fermion propagating in 
the five dimensional space.

Curiously enough, the tightest constraints in the case of $\tmf$
do not arise from truly CPT-violation oriented experiments.
%Curiously enough, there are not truly CPT-violation oriented experiments 
%that bring the tightest constraints in the case of $\tmf$. 
We exploit the Lorentz non-invariance of this background  
and estimate the level of sensitivity 
of the clock comparison experiments \cite{Amh} and \cite{Harvard1}
to the noncommutative scale as $(\tmf)^{-1/2} \sim 5\cdot 10^{11}$ GeV. 

An interesting feature of our result is the non-decou- pling of heavy KK 
modes due to the ultraviolet enhancement brought by $\theta_{MN}$. 
This behavior may change in the non-commutative SUSY theories which
phenomenological consequences 
obviously deserve more careful analysis. 

%The reason is that SUSY theories converge 
%much faster in the UV regime and therefore physical observables 
%could be free (or almost free) from the uncertainties 
%related to the choice of the cutoff. 

\bigskip {\bf Acknowledgments} M.P. thanks T. ter Veldhuis and D. Demir for 
interesting stimulating discussions. R.R. thanks W.Siegel for discussions. 
This work was supported in part by the Department of Energy under 
Grant No. DE-FG02-94ER40823 and NSF grant PHY-9722101.

\end{document}